\documentclass[aps,prl,showpacs,showkeywords,twocolumn,superscriptaddress]{revtex4-1}

\usepackage{graphicx, graphics}% Include figure files
\usepackage{mathcomp}
\usepackage{psfrag}
\usepackage{amsmath,amssymb,lmodern}
\usepackage{dcolumn}% Align table columns on decimal point
\usepackage{bm}% bold math
\usepackage[below]{placeins}
\usepackage{array}
\usepackage[usenames]{color}
\usepackage{textcomp}
\begin{document}

\title{Magnetic phase transitions of spin-1 ultracold bosons \\ in a cubic optical lattice}
\author{Yongqiang Li}
\affiliation{Department of Physics, National University of Defense Technology, Changsha 410073, P. R. China}
\author{Liang He}
\affiliation{Institute for Theoretical Physics, Technical University Dresden, D-01062 Dresden, Germany}
\author{Walter Hofstetter}
\affiliation{Institut f\"ur Theoretische Physik, Goethe-Universit\"at, 60438 Frankfurt/Main, Germany}
%\author{Jianmin Yuan}
%\affiliation{Department of Physics, National University of Defense Technology, Changsha 410073, P. R. China}
\date{\today}

\begin{abstract}
We investigate strongly correlated spin-1 ultracold bosons with antiferromagnetic interactions in a cubic optical lattice, based on bosonic dynamical mean-field theory. Rich phase diagrams of the system are mapped out at both zero and finite temperature, and in particular the existence of a spin-singlet condensate is established. Interestingly, at finite temperature, we find that the superfluid can be heated into a Mott insulator with even (odd) filling via a first- (second-) order phase transition, analogous to the Pomeranchuk effect in $^{3}$He. Moreover, for typical experimental setups, we estimate the critical temperatures for different ordered phases and our results suggest that direct experimental observation of these phases is promising.
\end{abstract}


\maketitle
Spin correlations play an important role in understanding ground states of spinor bosons~\cite{Lewenstein}, and may lead to collective phenomena that are absent in the scalar condensate~\cite{early_work_1}. In recent years, spinor ultracold bosons have been successfully realized in experiments~\cite{Ketterle_1998}, and widely studied including spin mixing~\cite{spin_1}, spin waves~\cite{spin_2}, spin dynamics~\cite{spin_3}, spin textures~\cite{texture} and phase transitions~\cite{L. Zhao}. Experimental realizations open a new path to studying quantum phenomena of spin-correlated condensates and competing insulating states, and also motivate theoretical studies on spinor Bose gases in an optical lattice~\cite{Demler, Yip, Imambekov, Snoek, Tsuchiya, Tsuchiya2, Mobarak, Batrouni, Natu}. While quantum Monte-Carlo simulations and DMRG were available in one spatial dimension~\cite{Batrouni}, up to now, these strongly correlated states in a three-dimensional (3D) optical lattice have been mainly investigated via lattice-gauge-theory predictions~\cite{Demler}, effective spin models in the strong-coupling limit~\cite{Imambekov,Snoek} or static mean-field theory~\cite{Tsuchiya,Tsuchiya2,Natu}. Beyond static mean-field theory, a complete phase diagram of spinor bosons in a 3D optical lattice for the full range from small to large couplings is still unknown. Further open issues are the existence of a spin-singlet condensate phase in a spinor bosonic system~\cite{Demler}, the stability of spin-ordered phases at finite temperature, and unconventional properties of quantum phase transitions arising from spin correlations~\cite{Lewenstein}.

\begin{figure*}[t]
\vspace{-15mm}
\hspace{-18mm}
\includegraphics*[width=1.0\linewidth]{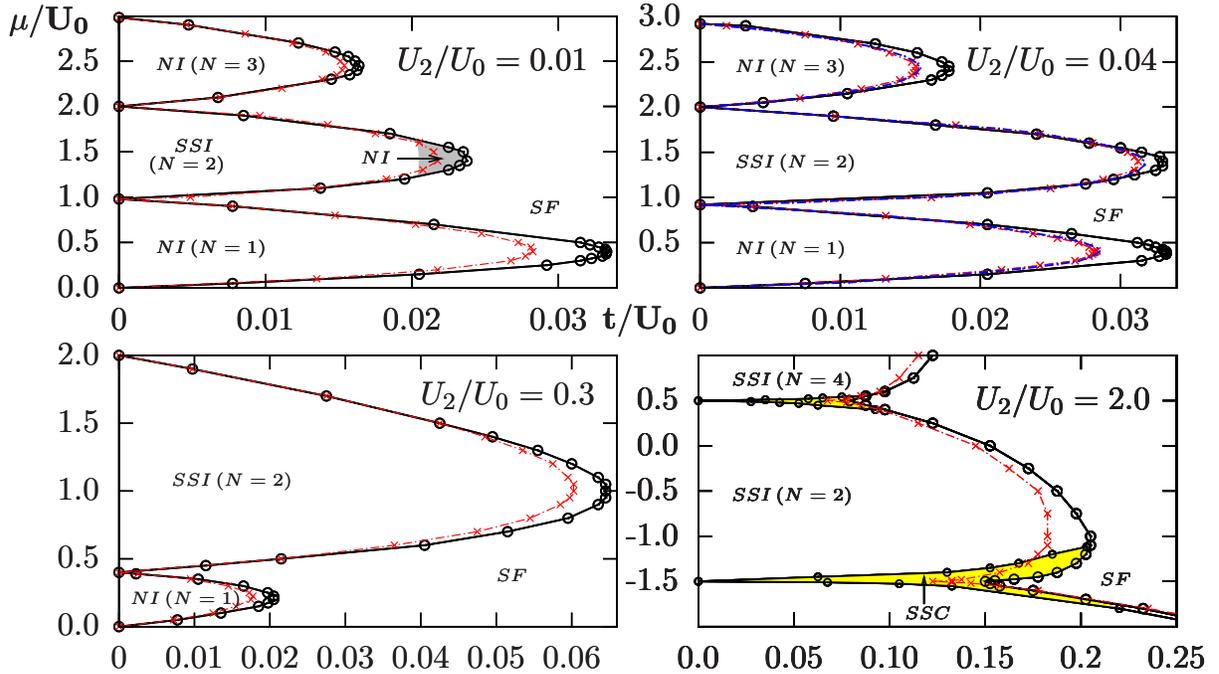}
\vspace{-5mm}
\caption{(Color online) Zero-temperature phase diagram for spin-1 ultracold bosons in a 3D cubic lattice for different anti-ferromagnetic interactions $U_2/U_0=0.01$, $0.04\;(\rm ^{23}Na)$, $0.3$, and $2.0$, respectively, obtained via BDMFT (black circle), Gutzwiller (red cross) and in Ref.~\cite{Tsuchiya2} (blue dashed). There are four different phases in these diagrams: superfluid (SF), nematic insulator (NI), spin-singlet insulator (SSI) and spin-singlet condensate (SSC). %In contrast to mean-field theories, these phases including SF and NI demonstrate non-zero local magnetization $M\neq0$, obtained by BDMFT. The Mott-insulating phase (MI) features filling-dependent spin structure: the MI with an odd number of atoms per site has spin $S\neq0$ (NI), whereas the MI with an even number of atoms shows competition between SSI and NI.
}\label{Ne_1}
\end{figure*}

\begin{figure*}[t]
\vspace{-7.5mm}
\begin{tabular}{ccc}
\hspace{-3mm}
\includegraphics[width=0.45\linewidth]{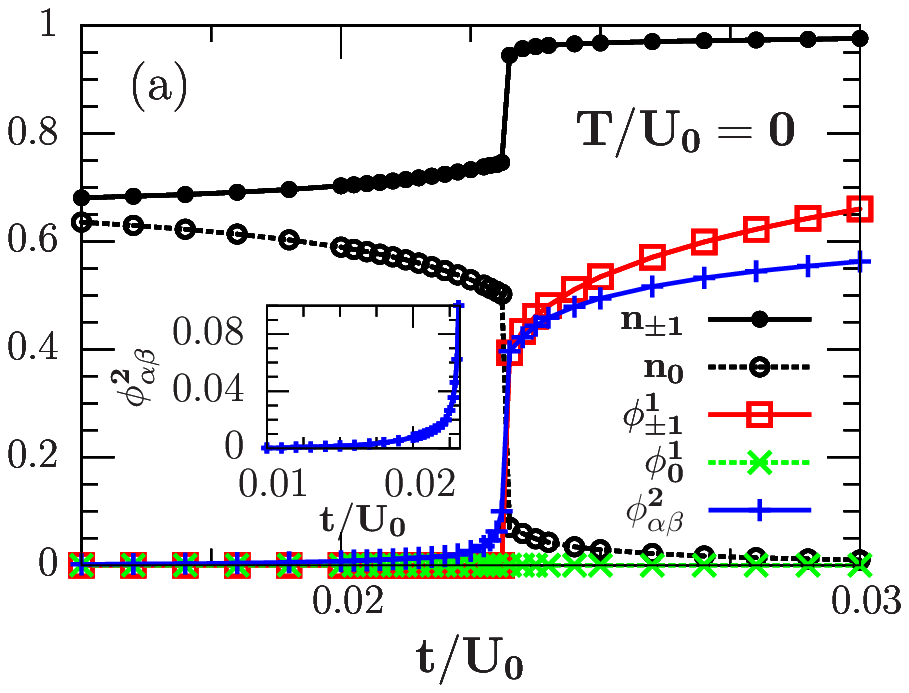}
\hspace{-23.mm}
\includegraphics[width=0.45\linewidth]{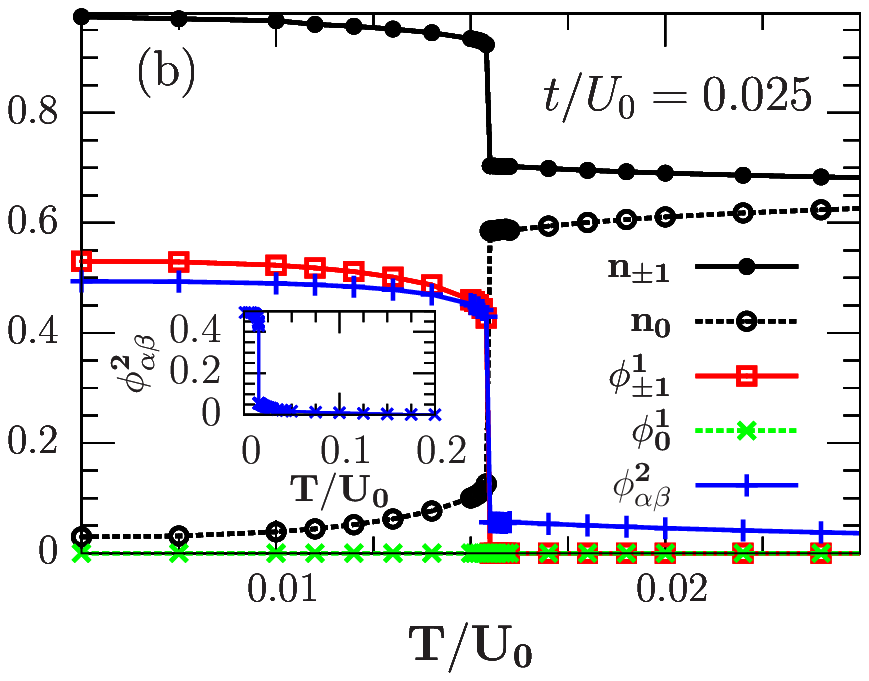}
\hspace{-25.mm}
\includegraphics[width=0.36\linewidth]{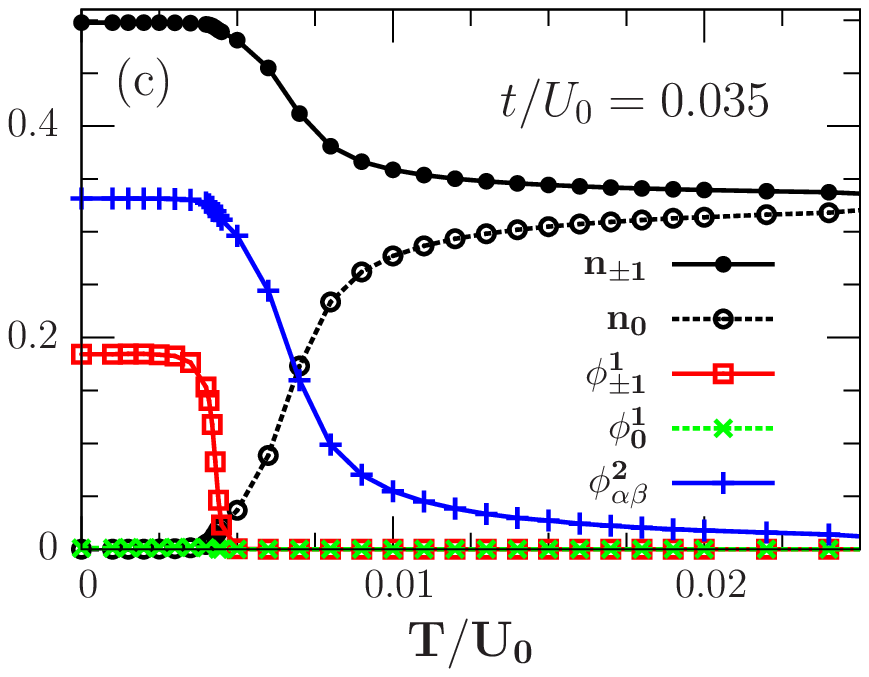}
\end{tabular}
\vspace{-3mm}
\caption{(Color online)  Zero-temperature SSI-NI-SF (a) and finite-temperature SF-NI-UI phase transitions (b)(c) in a 3D optical lattice with chemical potential $\mu/U_0=1.35$ (a)(b) and 0.35 (c), respectively, and interaction $U_2/U_0=0.01$ (upper left panel in Fig.~\ref{Ne_1}, and Fig.~\ref{density_trap}). Inset: zoom of the main figure around the critical point of nematic order (a), and zoom of the finite-temperature phase transition (b). BDMFT predicts a first- (second-) order superfluid-nematic-insulating phase transition for even (odd) fillings upon increasing temperature, analogous to the Pomeranchuk effect in $^3$He.}\label{order}
\end{figure*}

As a first step to bridge this gap, here we investigate properties of a strongly correlated spinor Bose gas in a 3D optical lattice at both zero and finite temperature by using a generalized bosonic dynamical mean-field theory~\cite{Byczuk2008,Hubener2009,Hu2009,Werner2010}. We establish a complete phase diagram of lattice spinor bosons with antiferromagnetic interactions, which is shown to contain superfluid (SF), spin-singlet condensate (SSC), nematic insulator (NI) and spin-singlet insulator (SSI) phases (see Fig.~\ref{Ne_1}). In particular, we establish the existence of a spin-singlet condensate. At finite temperature, we find a superfluid-nematic-insulating phase transition around the tip of Mott lobe upon heating, analogous to the Pomeranchuk effect in liquid $^3$He (see Fig.~\ref{order}). However, this transition is found to be first order for even fillings, but second order for odd fillings, contrary to spinless case.

In sufficiently deep lattices, spin-1 bosons with spin-symmetric tunneling are well described by a generalized Bose-Hubbard model~\cite{Lewenstein},
\begin{eqnarray}\label{Hamil}
\hat{H} &=& - t\sum_{\langle ij \rangle,\sigma}(b^\dagger_{i\sigma}b_{j\sigma} + b^\dagger_{j\sigma }b_{i\sigma})
           +\frac12U_0\sum_i n_i(n_i-1) \nonumber\\
        && +\frac12U_2\sum_i ({\bf S}^2_i - 2n_i) -\mu\sum_i n_i,
\end{eqnarray}
where $b^\dagger_{i\sigma}$ ($b_{i\sigma}$) is the bosonic creation (annihilation) operator of hyperfine state $m_F =\sigma$ at site $i$, $n_i=\sum_\sigma n_{i\sigma}$ with $n_{i\sigma}\equiv b^\dagger_{i\sigma}b_{i\sigma}$ being the number of particles in hyperfine state $\sigma$ at site $i$, ${\bf S}_i\equiv b^\dagger_{i\sigma}{\bf F}_{\sigma\sigma^\prime}b_{i\sigma^\prime}$ is the total spin operator at site $i$ with ${\bf F}_{\sigma\sigma^\prime}$ being the spin matrices for spin-1, $\mu$ denotes the chemical potential, and $t$ the hopping matrix element between nearest neighbours on the lattice. The second term in Eq.~(\ref{Hamil}) describes a Hubbard repulsion between atoms with $U_0=(g_0+2g_2)/3$, and the third term describes on-site spin-dependent interactions with $U_2=(g_2-g_0)/3$. Here $g_s=4\pi \hbar a_s /M_a $ with $a_s$ being the s-wave scattering length in the spin $s$ channel and $M_a$ the atomic mass. In the following we consider the case of antiferromagnetic interaction $g_0<g_2$. This is the case, for example, in experiments with $^{23}{\rm Na}$ atoms with $U_2/U_0\simeq0.04$~\cite{Burke,L. Zhao}.

In the limit with vanishing spin-dependent interaction $U_2$, the lattice model in Eq. (\ref{Hamil}) reduces to a set of independent Bose-Hubbard model, where a conventional superfluid-insulator transition occurs when $U_0/t$ is tuned across the transition point. On the contrary, in the limit with $U_2 \gg U_0$, spin-singlet pairs \big($\langle{\bf S}^2\rangle =0$\big) are energetically favorable for an even number of atoms in each well. In the intermediate, experimentally accessible regime with $0<U_2 \simeq U_0$~\cite{Burke,L. Zhao}, the system has an even richer phase diagram as a result of spin-dependent interactions, which compete both with quantum and thermal fluctuations resulting in different spin-correlated states. We will discuss this regime in more detail in the following.

We investigate the system by means of spinor bosonic dynamical mean-field theory (BDMFT), which allows us to investigate the system in an unified framework since it is non-perturbative and captures the local quantum fluctuations exactly. Indeed, BDMFT has been developed~\cite{Byczuk2008} and implemented~\cite{Hubener2009,Hu2009,Werner2010} successfully for the single- and two-component Bose-Hubbard models, where it provides a quantitative description for strongly correlated systems in a 3D optical lattice, and the validity of this approach has been verified against quantum Monte-Carlo simulations~\cite{QMC}. Inspired by this success, a spinor version of BDMFT is formulated here, where the physics on each lattice site is determined from a local effective action obtained by integrating out all other degrees of freedom in the lattice model Eq.~(\ref{Hamil}), excluding the lattice site considered. The local effective action is then represented by an Anderson impurity model, which is solved by exact diagonalization~\cite{Hubener2009}.
%Note that BDMFT takes local quantum correlations into account but neglects spatial dynamical correlations. The latter can be included within quantum Monte Carlo simulations or (to some degree) in cluster BDMFT.
%The Hamiltonian Eq.~(\ref{Hamil}) is mapped onto an effective single-site problem, and the physical properties on each lattice site are determined by a local effective action which is captured by an Anderson impurity model solved by exact diagonalization~\cite{Hubener2009}.
For comparison, a spinor bosonic Gutzwiller mean-field theory is implemented as well, and we find remarkable agreement with results from the Gutzwiller ansatz in Ref.~\cite{Tsuchiya2}, as shown in Fig.~\ref{Ne_1}.

\emph{Zero temperature ---} Our main results for the zero-temperature case are summarized in Fig.~\ref{Ne_1}, where a set of zero-temperature phase diagrams of spinor bosonic gases in a cubic optical lattice is shown for different anti-ferromagnetic interaction strengths $U_2/U_0=0.01$, $0.04\;(\rm ^{23}Na)$, $0.3$, and $2.0$. Four distinct phases, namely superfluid (SF), spin-singlet condensate (SSC), nematic insulator (NI) and spin-singlet insulator (SSI), are found. They are characterized according to the value for the condensate order parameter $\phi^1_\alpha\equiv\langle b_\alpha\rangle$, the nematic order parameter $\phi^2_{\alpha\beta} \equiv \langle S^\dagger_\alpha S_\beta \rangle -\delta_{\alpha\beta}/3 \langle S^2 \rangle$, the pair condensate order parameter $\varphi^2_{\alpha\beta}\equiv\langle b_\alpha b_\beta \rangle$, and the local magnetization $\bf {M}\equiv\langle {\bf S} \rangle$. We remark here that our results correctly recover unconventional spin ordering both in the atomic limit $U_0/J=\infty$ and in the weakly interacting regime. We discuss these phases in detail in the following.

In the deep Mott insulator (MI) with $U_{0},U_{2} \gg t$, we find that the system favors Mott insulating phases with different types of spin order in different interaction regimes, i.e. NI characterized by $\phi^1_\alpha =0 $, $\phi^2_{\alpha\beta}>0$ and $\bf {M}=0$; and SSI for an even number of atoms per site as a result of the formation of singlet pairs characterized by $\phi^1_\alpha =0$ and $\phi^2_{\alpha\beta}=0$ (due to the symmetry of the spin-wave function on each site, $n_i+S_i=({\rm even})$~\cite{Imambekov}, with $S_i$ being the total spin on site $i$). Actually, in the strong coupling limit $U_0,U_2 \gg t$, one can apply second-order perturbation theory and generate an effective spin Hamiltonian. Here, for filling $N\equiv\sum_{i}\langle n_{i}\rangle/N_{\mathrm{lat}}=({\rm odd})$ where $N_{\mathrm{lat}}$ is the number of lattice sites, it yields a spin model for $S\equiv\sum_{i} S_{i}/N_{\mathrm{lat}}=1$ bosons~\cite{Imambekov, Snoek} for describing the interplay between ferromagnetic and nematic long-range order,
\begin{eqnarray}
{\hat H_{eff}} = -J_1 \sum_{\langle ij\rangle} {\bf S}_i \cdot {\bf S_j} - J_2 \sum_{\langle ij\rangle}({\bf S}_i \cdot {\bf S_j})^2,
\end{eqnarray}
where $\langle ij\rangle$ denotes the nearest neighbor sites $i$, $j$. One can see that the $J_{1}$ term favors the ferromagnetic phase with $\mathrm{\mathbf{M}}\neq0$, but the $J_{2}$ term favors the nematic phase with $\phi_{\alpha\beta}^{2}>0$ and ${\bf {M}=0}$. The parameter choices $U_{2}/U_{0}=0.04$ and $U_{2}/U_{0}=0.3$ in Fig.~\ref{Ne_1} with a total filling $N=1$, $3$, for example, correspond to $J_{2}/J_{1}=1.5$, 1.9 and $J_{2}/J_{1}=3.3$, 3.8, respectively. We find that the system is in a nematic insulating phase in both cases, which is consistent with the quantum Monte-Carlo predictions for the above effective spin model~\cite{Harada}. Note that the three spin components are not equally populated in the nematic phases, while an equal mixture occurs in the static mean-field predictions~\cite{Lewenstein}. Similar findings were also reported in one dimensional case~\cite{Batrouni} and could be attributed to the quantum fluctuations neglected in the static mean-field approach.

Away from the deep MI regime, where the above results for $U_{0},U_{2} \gg t$ based on the perturbation theory are questionable, our simulations show that both NI and SSI phases are stable against quantum fluctuations within the Mott lobes, indicating a higher chance for directly
observing these phases in realistic experiments at finite temperature. For the SF-MI transition, we confirm that the transition is first order around the tip of the Mott lobe for even filling, while it is second order for odd filling in a 3D optical lattice~\cite{Lewenstein}, as shown in Fig.~\ref{order}(a). Note that, for small $U_2/U_0$ (upper-left panel in Fig.~\ref{Ne_1}), the interplay between tunneling and spin-dependent $U_2$ yields a SSI-NI phase transition before the superfluid-insulator transitions occurs~\cite{Imambekov}.
%BDMFT also captures the physics in the intermediate coupling regime (around the vicinity of the Mott transition), and predicts that insulating and strongly correlated superfluid phases have zero values of local magnetization $\bf {M}$ as well.

Interestingly, we find that upon increasing the spin-dependent interaction strength $U_{2}$, the lobe of SSI expands, while the Mott lobes for odd fillings shrink. Remarkably, we notice a SSC region, characterized by $\phi_{\alpha}^{1}=0$, $\phi_{\alpha\beta}^{2}=0$, $\langle{\bf S}^{2}\rangle=0$, ${\bf {M}=0}$ and $\Theta\equiv\langle-2b_{-1}b_{1}+b_{0}^{2}\rangle\neq0$, which emerges in the phase diagram between the SSI Mott lobes after the Mott lobes for odd fillings finally vanish when $U_{2}/U_{0}>0.5$. Here, we in fact numerically establish the existence of SSC for spinor bosons in a 3D optical lattice (see the right-bottom plot of Fig.~\ref{Ne_1}, where a pronounced region of SSC are found for large spin-dependent
interaction $U_{2}/U_{0}=2$), which was first predicted in Ref.~\cite{Demler}.

%Finally, in weakly interacting systems with $t\gg U_{0},U_{2}$, the expectation values of the spin components are zero along any direction~\cite{Tsuchiya,Lewenstein}, {\it i.e.} a polar ordering with $\bf {M} =0$, which is also predicted in our simulations. For relatively large hopping, we find a superfluid phase characterized by $\phi^1_\pm \neq 0$ ($\bf {M} =0$).
Finally, in the weakly interacting regime with $t\gg U_{0},U_{2}$, a polar phase with zero magnetization $\mathrm{\mathbf{M}}=0$ is found in our simulations, which is consistent with findings in previous works~\cite{Tsuchiya,Lewenstein}.

\begin{figure*}
\vspace{-7.5mm}
\begin{tabular}{cc}
\hspace{-4mm}
\includegraphics*[width=0.52\linewidth]{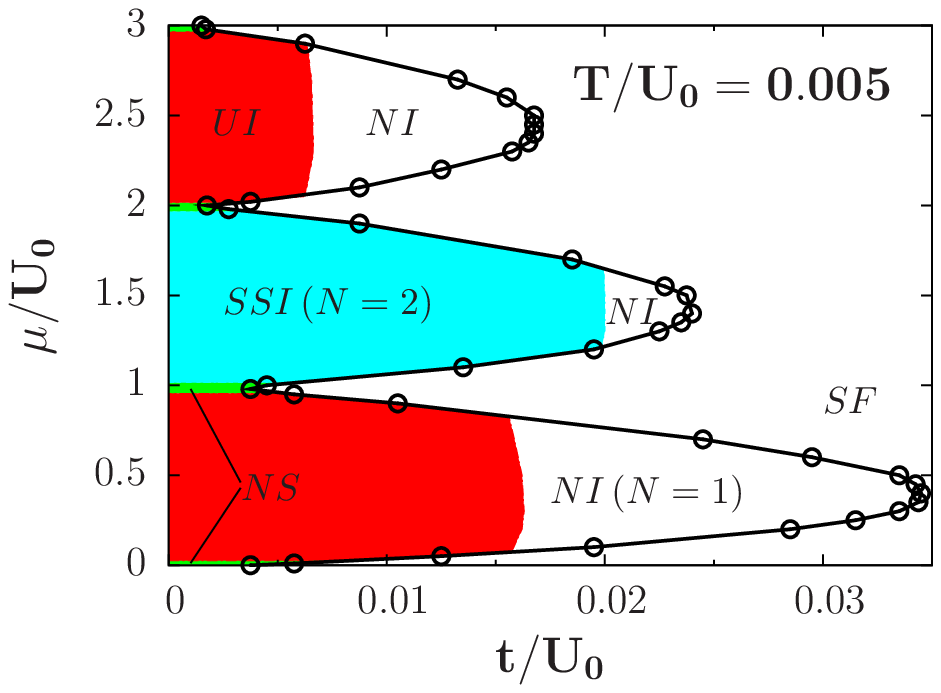}
\hspace{-6.5mm}
\includegraphics*[width=0.52\linewidth]{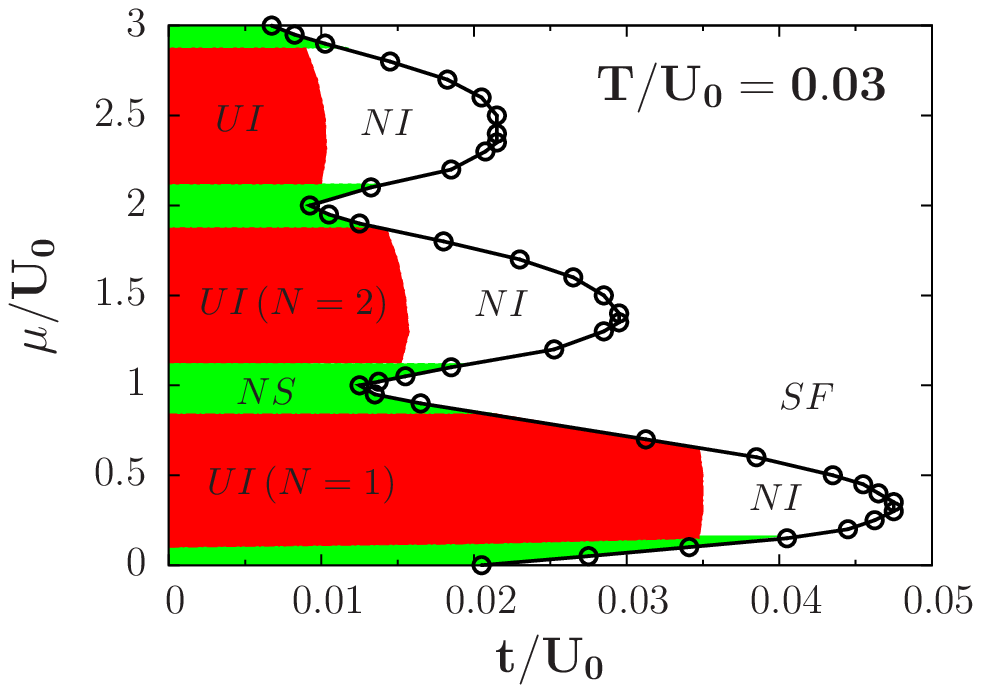}
\end{tabular}
\vspace{-3mm}
\caption{(Color online) Finite-temperature phase diagram for spin-1 ultracold bosons in a 3D cubic lattice with an anti-ferromagnetic interaction $U_2/U_0=0.01$ for different temperatures $T/U_0=0.005$ (left) and $T/U_0=0.03$ (right). There are five different phases in this diagram: superfluid (SF), nematic insulator (NI), spin-singlet insulator (SSI), unordered insulator (UI) and normal state (NS). SSI is fragile against temperature and shrinks into NI with increasing temperature, while UI develops from the lower hopping regions of NI, due to thermal fluctuations.}\label{density_trap}
\end{figure*}

\emph{Finite temperature ---} One crucial question regarding direct observation of the above spin correlated phases in realistic experiments is their stability against thermal fluctuations. For a typical case with $U_2/U_0 = 0.01$, the influence of thermal fluctuations is shown in Fig.~\ref{density_trap} in
terms of phase diagrams at finite temperatures $T/U_0=0.005$ (left panel) and $T/U_0=0.03$ (right panel), where five different phases appear in the system, including SF, NI, SSI, unordered Mott insulator (UI), and normal state (NS) which exists in the low hopping region between the insulating
lobes and is characterized by $\phi^1_\alpha=0$ but with a non-integer filling, as shown by the green filled region in Fig.~\ref{density_trap}. We remark here that these choices of temperature are within reach of present cooling schemes, such as spin-gradient cooling~\cite{Ketterle} and the coexistence of these phases can be achieved via an external harmonic trap.

We observe that the SF remains robust against small finite temperature, with a tiny change of the insulator-superfluid boundary for $T/U_0=0.005$. However, for higher temperature $T/U_0=0.03$, a large shift of the MI-SF boundary towards larger hopping is observed.
In addition, upon increasing temperature, we observe a transition from SF to NI (see Fig.~\ref{order}(b)), analogous to the Pomeranchuk effect in liquid $^{3}$He where the system favors localization upon heating at low temperature, since spin fluctuations in the Mott insulating phase can carry more entropy than the superfluid. Upon further increasing temperature, a NI to UI transition is observed.
Interestingly, we find that the superfluid-Mott-insulating phase transition for even fillings is first order, as shown in Fig.~\ref{order}(b). The physical reason is that the Mott state with even filling prefers the lowest spin state (spin $S=0$), while the SF contains higher spin states, similar to the zero-temperature case [Fig.~\ref{order}(a)].

For the SSI phase ($N=2$ with $U_2/U_0=0.01$), we find that it is sensitive to finite temperature, and the underlying physics is that the spin-gap $U_2$ term is small and easily destroyed by thermal fluctuations. As shown in the right panel of Fig.~\ref{density_trap}, for example, no SSI occurs at temperature $T/U_0=0.03$.
For experiments with $^{23}{\rm Na}$ ($U_{2}/U_{0}\approx0.04$) in a 3D cubic lattice formed by laser beams of wavelength 1064 nm and intensity $V_I\approx16\, E_{R}$, we find the critical temperature to be $T_c\approx1\, {\rm nK}$, where $E_{R}$ denotes the recoil energy. In addition, the nematic phase, NI (the critical temperature is $T_c\approx 5$ nK, for the same experimental setup with $^{23}{\rm Na}$ but with an intensity of $V_I\approx14\, E_{R}$), is reduced with increasing temperature, in favor of developing a non-ordered Mott state. We remark here that these correlated insulating states can be detected from their excitation spectra or density correlations~\cite{Imambekov, Natu}, by using, for instance, Bragg scattering~\cite{Imambekov}, quantum gas microscopy~\cite{single_site} or optical birefringence~\cite{Natu}, as long as the lifetimes of these states of spinor lattice bosons are sufficiently long~\cite{L. Zhao}. In recent experiments, a long lifetime steady state of spinor bosons in optical lattices was reported~\cite{L. Zhao}, as well as the existence of spin-nematic ordering in a spherical trap~\cite{T. Zibold}.

In conclusion, we have investigated quantum phases of ultracold spinor Bose gases loaded into a cubic optical lattice, based on spinor bosonic DMFT.
We obtain the complete phase diagram of spinor bosons with antiferromagnetic interactions in an optical lattice at both zero and finite temperature. In particular, we demonstrate the existence of a spin-singlet condensate. Interestingly, we observe that the superfluid can be $\it heated$ into a Mott insulator with even (odd) filling via a first- (second-) order phase transition, analogous to the Pomeranchuk effect in $^3$He. We find that the critical temperature of ordered states, such as nematic phase and spin-singlet insulator, is within reach of present cooling schemes, indicating the chance to directly observe these phases using current experimental techniques.

We acknowledge useful discussions with J.-M. Yuan and Z.-X. Zhao. This work was supported by the National Natural Science
Foundation of China under Grants No. 11304386 and No. 11104350, and by the Deutsche Forschungsgemeinschaft DFG via Sonderforschungsbereich SFB-TR/49, and by ZUK 64 (L.H.).

\FloatBarrier

\end{document}